\documentclass[11pt]{article}

\usepackage{mathtools}
\usepackage{slashed}
\usepackage{xcolor} % Define colors
\usepackage{graphicx}
\usepackage{dcolumn} % Align table columns on decimal point
\usepackage{bm} % Bold math
\usepackage{epsfig}
\usepackage{epstopdf}
\usepackage{grffile}
\usepackage{color}
\usepackage{colordvi}
\usepackage{amsmath,amssymb}
\usepackage{rotating}
\usepackage{lscape}
\usepackage{cite}
\usepackage{float}
\usepackage{hyperref}
\usepackage{subfigure}

\begin{document}
\begin{center}
\vspace*{15mm}
\vspace{1cm}
{\Large \bf Probing Higgs boson couplings in $t\bar{t}b\bar{b}$ production at the LHC}
\vspace{1cm}

Hoda Hesari
 \vspace*{0.5cm}

School of Particles and Accelerators, Institute for Research in Fundamental Sciences (IPM) P.O. Box 19395-5531, Tehran, Iran  \\

\vspace*{.2cm}
\end{center}

\vspace*{10mm}

%
%%%%%%%%%%%%%%%%%%%%%%%%%%%%%%%%%%%%%%%%%%    abstract    %%%%%%%%%%%%%%%%%%%%%%%%%%%%%%%%%%%%%%%%%%%%%%%%%
%
\begin{abstract}\label{abstract}
In this paper, we investigate the effects of dimension six operators in the framework of the standard model (SM) effective field theory on using $t\bar{t}b\bar{b}$ channel at the LHC with the center-of-mass energy of 13 TeV. In this analysis, we only assume leptonic top-quarks. Once the events were simulated, the CMS-measured cross-section was used to constrain the corresponding effective Wilson coefficients at $95\%$ confidence level. Considering the SM effective Lagrangian, the $t\bar{t}b\bar{b}$ events in proton-proton collision were very sensitive to the effective Wilson coefficients $ \bar{C}_{g},\tilde{C_{g}},\bar{C}_{uG},\bar{C}_{uW}$. We show that using $t\bar{t}b\bar{b}$ events, we can provide stronger constraints on these Wilson coefficients than the current ones obtained from other channels.  
  
\end{abstract}

\newpage

%%%%%%%%%%%%%%%%%%%%%%%%%%%%%%%%%%%%%%%%%%    Introduction    %%%%%%%%%%%%%%%%%%%%%%%%%%%%%%%%%%%%%%%%%%%%%%%%%
\section{Introduction}\label{sec:intro}
\begin{figure}[t!]
%\begin{center}
\subfigure{\includegraphics[height=1.7cm,width=0.17\textwidth]{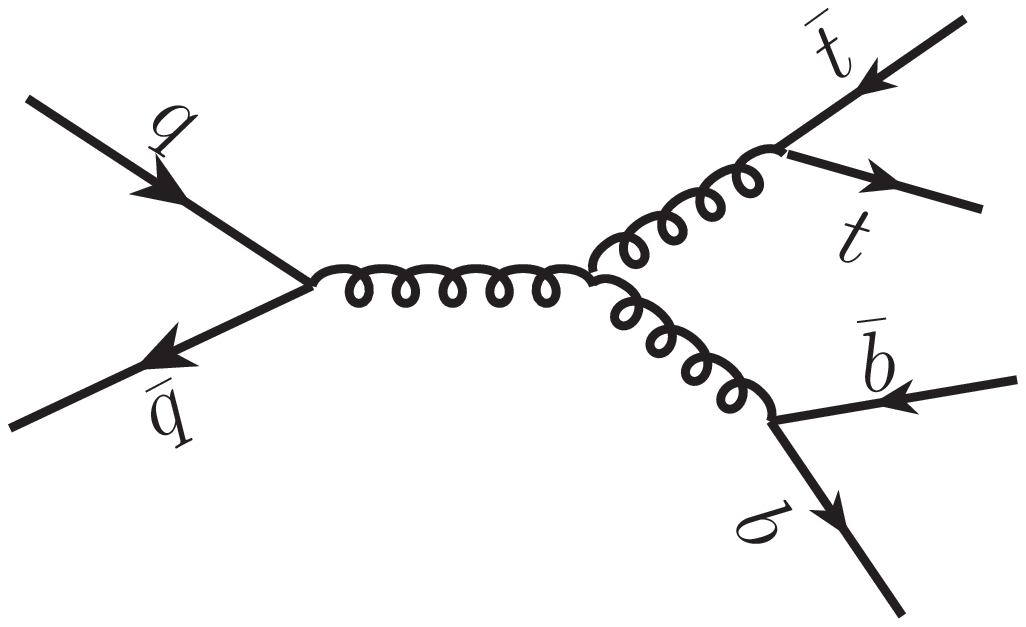}}
   \subfigure{\includegraphics[height=1.7cm,width=0.17\textwidth]{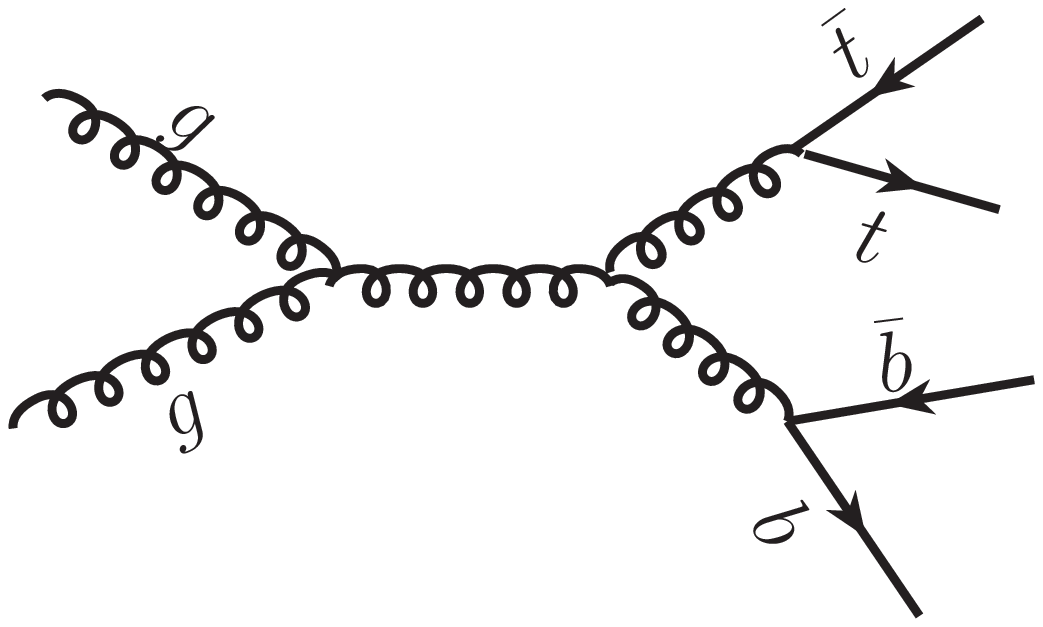}}
    \subfigure{\includegraphics[height=1.7cm,width=0.17\textwidth]{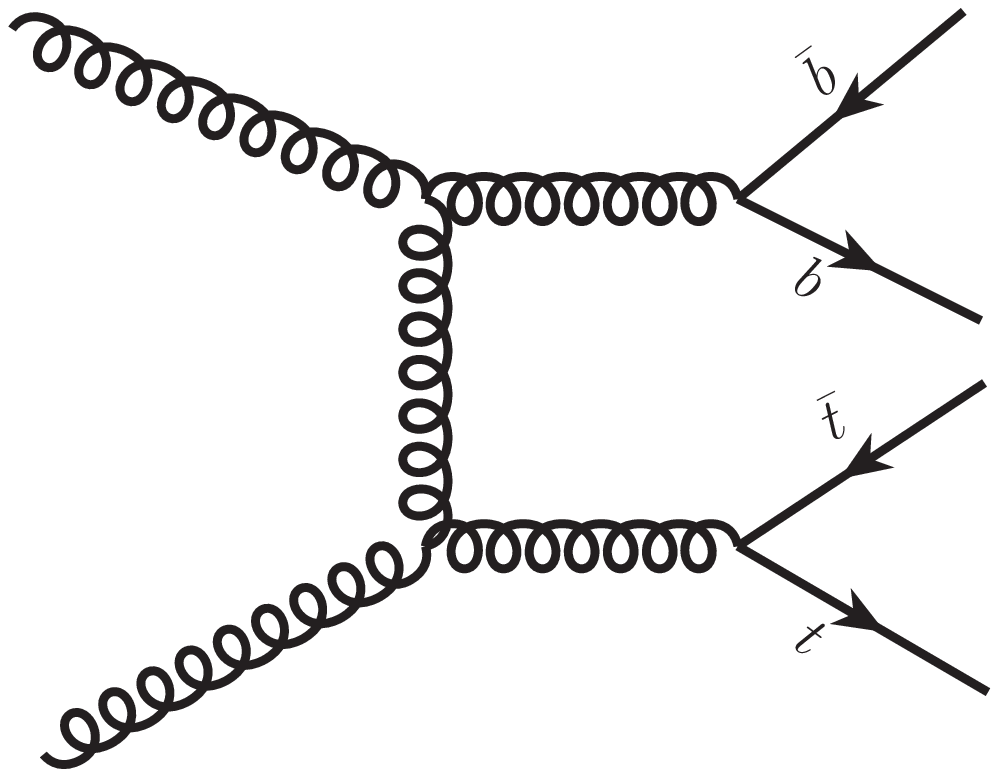}}
     \subfigure{\includegraphics[height=2.2cm,width=0.16\textwidth]{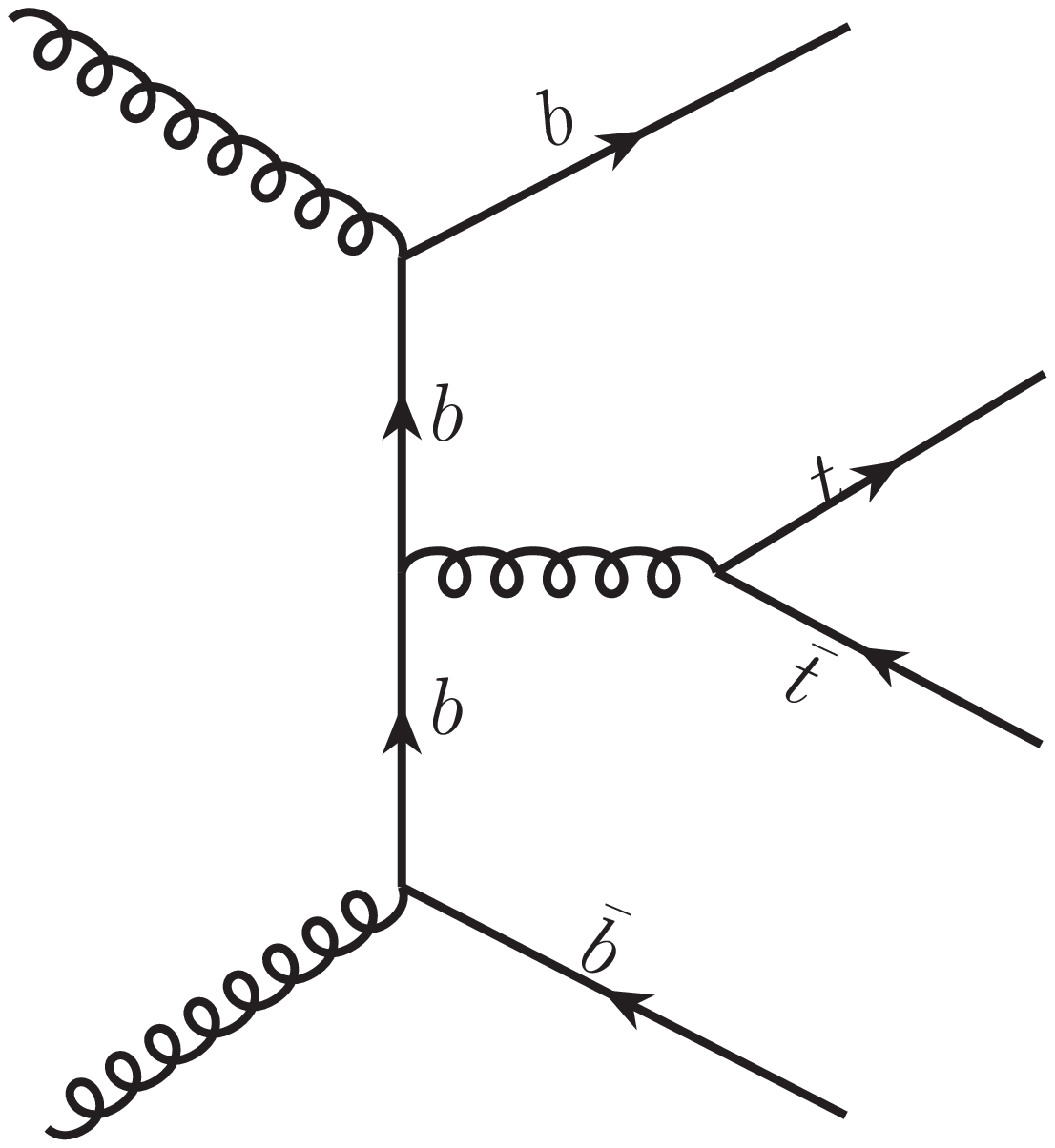}}
     \subfigure{\includegraphics[height=2.2cm,width=0.16\textwidth]{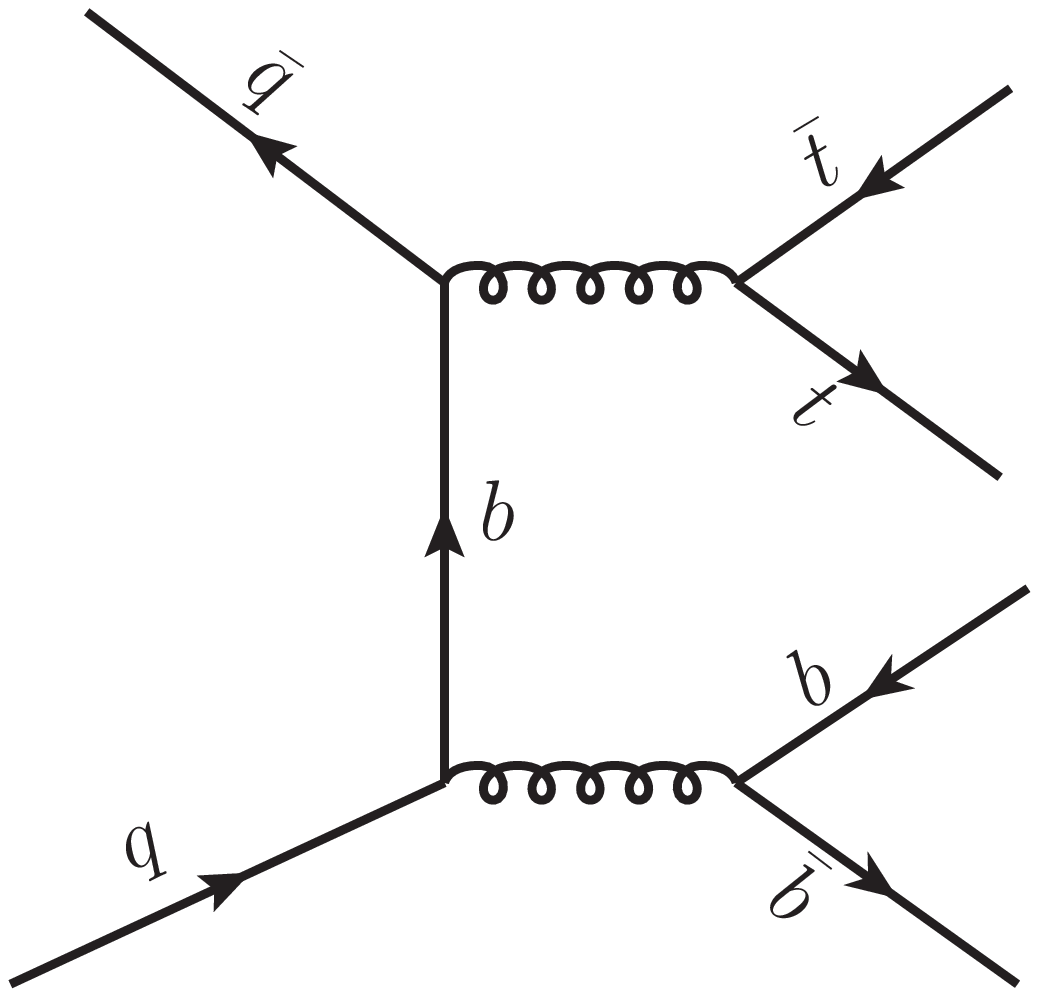}}
%#\end{center}
\caption{ Sample Feynman diagrams with tree order for $q \bar{q}\rightarrow t\bar{t}b\bar{b} $ and $ g g \rightarrow t\bar{t}b\bar{b} $ events production. }\label{ttbb-LO}
\end{figure}

Third generation of quarks ( top and bottom quarks), because of their relatively larger masses have been an attractive final states in many studies. Among the various combinations of the third generation quarks, $t\bar{t}b\bar{b}$ is of particular merit. There are two main reasons for such consideration. First of all, this channel is particularly important for Higgs studies. After the discovery of the Higgs boson, studying the interactions of this particle with other particles such as heavy fermions like top quarks are the next goals of detection in CMS and ATLAS in the LHC experiment\cite{atlash,cmsh}. A method of studying the Higgs boson interactions is to investigate the events of Higgs particle production associated with top pairs($t\bar{t}H$) and the Higgs boson is most likely decay to $b \bar{b}$, thus the final event is $t\bar{t}b\bar{b}$ and also the $t\bar{t}b\bar{b}$ events are considered as the irreducible background events of the $t\bar{t}H$ events.  For example, in \cite{Jo:2015zxa}, the two observables $m_{b\bar{b}}$ and $\Delta R_{b\bar{b}}$ were proposed for separating $t\bar{t}H$ from $t\bar{t}b\bar{b}$. \\
Another place that this channel may become important is in hunting new particles, especially higgs-like scalars. In some models such as 2HDM in addition to the SM Higgs, another Higgs is predicted. These new Higgs usually decay to heavy quark particles; thus, most studies have focused on the final states of $t\bar{t}b\bar{b}$\cite{Gori:2016zto,Hajer:2015gka,Pomarol:2008bh}. Some papers have addressed the $t\bar{t}X$ and $b\bar{b}Y$ events in simplified models, \cite{Alvarez:2017wwr,Dolan:2016qvg}, wherein X and Y represent the newly produced particles. As soon as these particles are decayed to a bottom quark pair or top quark pair, the final state of these events can be $t\bar{t}b\bar{b}$,$t\bar{t}t\bar{t}$ ,$b\bar{b}b\bar{b}$. In \cite{Alvarez:2017wwr}, the observables $M_{b\bar{b}},P_{T}^{b}$ were used to study new physics through the $t\bar{t}b\bar{b}$ channel. Generally, the $t\bar{t}b\bar{b}$ channel better suits studying new physics than the two other channels, namely $t\bar{t}t\bar{t}$ and $b\bar{b}b\bar{b}$, since due to less populated state of the $t\bar{t}b\bar{b}$ channel than the $t\bar{t}t\bar{t}$ events, they are simpler to be reconstructed. Also, the $t\bar{t}b\bar{b}$ events in the presence of leptons have less QCD backgrounds than the $b\bar{b}b\bar{b}$ events. In \cite{Dolan:2016qvg}, it was shown that, using the kinematic observables like $M_{b\bar{b}},P_{T}^{b}$, the quantum numbers, such as spin and parity, of the new particles can be determined. As for the center-of-mass energy of 13TeV, the ATLAS group studied the new physics in VLQ and 2HDM models for the final state of $t\bar{t}b\bar{b}$ final state \cite{ATLAS:2016btu}. ATLAS group studied the 
final state of $b\bar{b}$ associated with heavy particle Y$(\rightarrow t \bar{t})$ in proton-proton collision at center of mass energy 13TeV, the results of which had many statistical uncertainties due to the less number of reconstructed particles. However, such a problem can be resolved by increasing the number of events.  \\
Because of the importance of this channel, it was studied extensively by CMS and ATLAS experiments at the center-of-mass energies of 7, 8, and 13TeV , followed by measuring the cross-section of the events. First, the CMS group observed the $t
\bar{t}b\bar{b}$ events at the LHC at the energy of 7TeV with 4.7fb$^{-1}$ data and precision of 3$\sigma$ relative to hypothesis zero. Besides, the fiducial cross-section was measured with 2$\sigma$ deviation from the SM \cite{Aad:2013tua}. The ATLAS group reported the fiducial cross-section in proton-proton collision with the energy of 8TeV and 20.3 fb$^{-1}$ data. The measured cross-section exhibited a deviation of nearly $1\sigma$ from the SM prediction. Moreover, they concluded that different models of gluon splitting ($g\rightarrow b \bar{b}$) could considerably affect the $t\bar{t}b\bar{b}$ cross-section \cite{Aad:2015yja}.
The CMS group studied the $t\bar{t}b\bar{b}$ events at the center-of-mass energy of 8TeV with 19.6 fb$^{-1}$ data at the LHC.
% In order to reduce the systematic error, the cross-sectional ratio $\frac{\sigma_{t\bar{t}b\bar{b}}}{\sigma_{t\bar{t}jj}}$ was reported. Accordingly, this ratio was observed  to have a 1.6$\sigma$ deviation from SM prediction. The cross-section measurement values of the events $\sigma_{t\bar{t}b\bar{b}}$ and $\sigma_{t\bar{t}jj}$ showed 2.5$\sigma$ and 1.5$\sigma$ deviations from SM, respectively \cite{CMS:2014yxa}. 
The cross-section measurement values of the events $\sigma_{t\bar{t}b\bar{b}}$ showed 2.5$\sigma$ deviations from SM \cite{CMS:2014yxa}. In the next work of the CMS group on $t\bar{t}b\bar{b}$ events at the center-of-mass energy of 8TeV, the CMS group studied the behavior of the $t\bar{t}b\bar{b}$ events with respect to some kinematic variables, such as $\Delta R_{b\bar{b}},M_{b\bar{b}},P_{T}^{b},\eta_{b}$ by using an integrated luminosity of 19.7fb$^{-1}$. The $t\bar{t}b\bar{b}$ events in some regions were observed to have some deviations from the predicted results in the SM
%however, the $ t\bar{t}jj$ events exhibited no such deviation on the observables. Furthermore
, they concluded that the b-tagging efficiency had the greatest effect on the measurement error of the cross-section of these events \cite{Khachatryan:2015mva}. In \cite{CMS:2016tlo}, the CMS group used 2.3fb$^{-1}$ at the center-of-mass energy of 13TeV to measure the cross-sections $\sigma_{t\bar{t}b\bar{b}}$ 
%, $\sigma_{t\bar{t}jj}$ as well as their ratio. The measured values were,
The measured values was
% respectively, 
equal to $176\pm5\pm33$ pb,
%, $3.9\pm5\pm3$ pb, and $0.022\pm0.003\pm0.006$ 
throughout the full phase space, while the predicted values up to the single-loop NLO order for the full phase space in the SM 
%were 
was equal to $257\pm26$ pb.
%, $3.2\pm0.4$ pb, and $0.012\pm0.001$ pb, respectively. 
Such a difference between the measured and predicted values can be a good topic for investigating new physics. The increased collision energy of the protons and the greater amount of data in RUN-2 of the LHC collider provide the ground to do a better investigation of the new interactions beyond the SM as well as the production of new particles.\\
To date, there has been no compelling direct  evidence indicating the contribution of new physics to recent measurements aimed to study physics beyond the SM. Thus, the new physics seems to have a scale well separated from the electroweak scale. Furthermore, there are numerous new models with new particles, in which the heavy degrees of freedom can be integrated out and removed. Such evidence prompts the use of the model-independent effective theory approach, in where the validity scale of the effective theory is less than the lightest particle produced in new physics. Thus far, numerous studies have been conducted to constrain the effective coefficients using the data derived from LHC\cite{Englert:2015hrx,Ferreira:2016jea,Sirunyan:2017uzs}. The present paper is aimed to investigate new physics in effective Lagrangian theory using the most recent results on the 
$t\bar{t}b\bar{b}$ events production cross-section at 
the LHC at the center-of-mass energy of 13TeV. \\
The $t\bar{t}b\bar{b}$ processes in the SM were performed via $q \bar{q}\rightarrow t\bar{t}b\bar{b} $ and $ g g \rightarrow t\bar{t}b\bar{b} $ processes. The number of Feynman diagrams for $ g g \rightarrow t\bar{t}b\bar{b} $ at tree level and $q \bar{q}\rightarrow t\bar{t}b\bar{b} $ is equal to 7 and 36, 
respectively. Some examples of these diagrams can be found in Figure (\ref{ttbb-LO}) \cite{Bredenstein:2010rs}. Being calculated at the single-loop order, the 
number of Feynman diagrams $q \bar{q}\rightarrow t\bar{t}b\bar{b} $ and $ g g \rightarrow t\bar{t}b\bar{b}$ mounted to 118 and 1003, respectively 
\cite{Bredenstein:2010rs}. The K-factor was calculated as about 1.8. However, by applying cuts on different observables such as $M_{b\bar{b}},P_{T}^{b}$, the K-
factor could be reduced to 1.24 \cite{Bredenstein:2009aj,Bevilacqua:2009zn}.\\ 
Section (\ref{sec:framework}) provides a brief description of the theoretical framework. Section (\ref{sec:analysis}) explains the $t\bar{t}b\bar{b}$ event simulation and analysis methods, besides investigating the corresponding Wilson coefficients as well as their effects on the $t\bar{t}b\bar{b}$ events production. Section (\ref{sec:results}) provides the results obtained from this approach and compares them with those of other studies. We summarize our conclusions in Section \ref{summ}. 

%
%%%%%%%%%%%%%%%%%%%%%%%%%%%%%%%%%%%%%%%%%%%%%%%%%%%%%%%%%%%%%%%%%%%%%%
\section{Theoretical framework}\label{sec:framework}
%%%%%%%%%%%%%%%%%%%%%%%%%%%%%%%%%%%%%%%%%%%%%%%%%%%%%%%%%%%%%%%%%%%%%%
%
Effective Lagrangian includes terms with mass dimensions higher than four. Dimension of these terms can be reduced using reverse powers of the new physics scale $\Lambda$. In this approach, all heavy new degrees of freedom can be integrated out\cite{Fuentes-Martin:2016uol}. In other words, the estimates made by effective operators would be valid until the effective physics scale is smaller than that of new physics. Considering all the SM gauge symmetries as well as the lepton and baryon numbers conservation, the first terms that can be obtained are those with the mass dimension of six, which will influence the observables more significantly than the higher-order terms. Totally, there are 59 six-dimension independent operators \cite{Grzadkowski:2010es}. The overall form of the effective Lagrangian is as follows\cite{Grzadkowski:2010es,Buchmuller:1985jz ,Hagiwara:1993ck}:
\begin{eqnarray}
\mathcal{L}_{eff} = \mathcal{L}_{SM} + \sum_{i}\frac{c_{i}O_{i}}{\Lambda^{2}},
\end{eqnarray}
where $O_{i}$ are independent operators that can be obtained in various bases. 
The present paper is focused on the SILH (strongly interacting light Higgs) basis. If it is supposed to concentrate on a particular model, the Wilson coefficients,$ c_{i}$s, can be estimated through matching computations. In the present paper, only the presence of SM particles was taken into consideration. Effective Lagrangian in the SILH basis is as \cite{PAlloul:2013naa}:
\begin{eqnarray}\label{leff1}
\mathcal {L} = \mathcal{L}_{SM} + \mathcal{L}_{SILH} + \mathcal {L}_{CP}
   + \mathcal{L}_{F_1} + \mathcal{L}_{F_2}  +  \mathcal{L}_{G}
\end{eqnarray}
Where $ \mathcal{L}_{SILH} $ represents the terms including Higgs field $(\Phi)$, in which CP is conserved and it has the following form \cite{PAlloul:2013naa}:
%------------------------------
\begin{eqnarray}\label{lSilh1}
	\begin{split}
		{\cal L}_{SILH} = & \ 
		\frac{g'^2 \bar{c}_{\gamma}}{m_{W}^2} \Phi^\dag \Phi B_{\mu\nu} B^{\mu\nu}     
		+ \frac{i g' \bar{c}_{B}}{2 m_{W}^2} [\Phi^\dag \overleftrightarrow{D}^\mu \Phi ] \partial^\nu  B_{\mu \nu} 
		+\frac{g_s^2 \bar{c}_{g}}{m_{W}^2} \Phi^\dag \Phi G_{\mu\nu}^a G_a^{\mu\nu} \\
		& \
		+\frac{\bar{c}_{u}}{v^2} y_u \Phi^\dag \Phi\ \Phi^\dag {\bar Q}_L u_R 
		+ \frac{\bar{c}_{d}}{v^2} y_d \Phi^\dag \Phi\ \Phi {\bar Q}_L d_R
		- [\frac{\bar{c}_{l}}{v^2} y_\ell\ \Phi^\dag \Phi\ \Phi {\bar L}_L e_R
		+ {\rm h.c.} ]  \\
		& \
		+\frac{i g \bar{c}_{W}}{2m_{W}^2} [ \Phi^\dag \sigma_{k} \overleftrightarrow{D}^\mu \Phi ]  D^\nu  W^k_{\mu \nu}
		+ \frac{ i g \bar {c}_{HW}}{m_{W}^2} \big[D^\mu \Phi^\dag \sigma_{k} D^\nu \Phi\big] W_{\mu \nu}^k  - \frac{\bar{c}_{6} \lambda}{v^2} \big[\Phi^\dag \Phi \big]^3 \\
		& \
		 + \frac{i g'\ \bar{c}_{HB}}{m_{W}^2}  [D^\mu \Phi^\dag D^\nu \Phi] B_{\mu \nu}   
		+ \frac{\bar{c}_{T}}{2 v^2} [ \Phi^\dag {\overleftrightarrow{D}}^\mu \Phi] [\Phi^\dag {\overleftrightarrow{D}}_\mu \Phi ]
		+ \frac{\bar{c}_{H}}{2 v^2} \partial^\mu[\Phi^\dag \Phi] \partial_\mu [ \Phi^\dagger \Phi ] 
	\end{split}
\end{eqnarray}
$\Phi^\dag \overleftrightarrow{D}^\mu \Phi $ indicates the Hermitian operator as $\Phi^\dag (D^{\mu}\Phi) - (D^{\mu}\Phi)^{\dag}\Phi$, $T_{2k}$ indicates matrices as half of the Pauli matrices $(T_{2k} = \sigma_k/2)$,  and $\mathcal{L}_{CP}$ represents those terms for which CP is violated and are represented as \cite{PAlloul:2013naa}:
\begin{eqnarray}\label{lcp1}
	  \mathcal{L}_{CP} = \frac{i g'\tilde c_{HB}}{m_W^2} D^\mu \Phi^\dag D^\nu \Phi {\widetilde B}_{\mu \nu} + \frac{g'^2\tilde{c}_{\gamma}}{m_W^2} \Phi^\dag \Phi B_{\mu\nu} {\widetilde B}^{\mu\nu} + \frac{i g \tilde{c}_{HW}}{m_W^2}  D^\mu \Phi^\dag T_{2k} D^\nu \Phi {\widetilde W}_{\mu \nu}^k +\\
  \frac{g^3 \tilde{c}_{3W}}{m_W^2} \epsilon_{ijk} W_{\mu\nu}^i W^\nu{}^j_\rho {\widetilde W}^{\rho\mu k}+\frac{g_s^2\tilde{c}_{g}}{m_W^2} \Phi^\dag \Phi G_{\mu\nu}^a {\widetilde G}^{\mu\nu}_a 
  +\frac{g_s^3 \tilde{c}_{3G}}{m_W^2} f_{abc} G_{\mu\nu}^a G^\nu{}^b_\rho {\widetilde G}^{\rho\mu c} 
       \end{eqnarray}
In $\mathcal{L}_{CP}$, the dual field tensors are defined as 
\begin{eqnarray}
   \widetilde G_{\mu\nu}^a = \frac12 \epsilon_{\mu\nu\rho\sigma} G^{\rho\sigma a},
   \widetilde W_{\mu\nu}^k = \frac12 \epsilon_{\mu\nu\rho\sigma} W^{\rho\sigma k},
  \widetilde B_{\mu\nu} = \frac12 \epsilon_{\mu\nu\rho\sigma} B^{\rho\sigma} .
\end{eqnarray}
The third term is $\mathcal{L}_{F_1}$. This term includes the interaction between two Higgs fields and a lepton or quark pair, in which :
\begin{eqnarray}\label{lf1}
	\begin{split}
\mathcal{L}_{F_1} =    &\
 + \frac{i \bar{c}_{Hd}}{v^2} [\bar d_R \gamma^\mu d_R]  [ \Phi^\dag{\overleftrightarrow D}_\mu \Phi]  + \frac{i \bar{c}_{Hu}}{v^2} [\bar{u}_R \gamma^\mu u_R] [\Phi^\dag{\overleftrightarrow D}_\mu \Phi]\\
   &\ 
   + \frac{i \bar{c}_{HL}}{v^2}  [\bar{L}_L \gamma^\mu L_L] [ \Phi^\dag{\overleftrightarrow D}_\mu \Phi] + \frac{4 i \bar{c'}_{HL}}{v^2} [\bar{L}_L \gamma^\mu T_{2k} L_L] [\Phi^\dag T^k_2 {\overleftrightarrow D}_\mu \Phi]\\
    &\ 
  +\frac{i \bar{c}_{He}}{v^2} [\bar{e}_R \gamma^\mu e_R][ \Phi^\dag{\overleftrightarrow D}_\mu \Phi] -[\frac{i \bar{c}_{Hud}}{v^2} [\bar{u}_R \gamma^\mu d_R] [ \Phi {\overleftrightarrow D}_\mu \Phi] + {\rm h.c.} ]   \\
  &\ 
   + \frac{4 i \bar{c'}_{HQ}}{v^2} [\bar{Q}_L \gamma^\mu T_{2k} Q_L] [\Phi^\dag T^k_2 {\overleftrightarrow D}_\mu \Phi] +\frac{i \bar{c}_{HQ}}{v^2}[\bar {Q}_L \gamma^\mu Q_L][ \Phi^\dag{\overleftrightarrow D}_\mu \Phi] \\
         \end{split}
\end{eqnarray}
The fourth terms of this Lagrangian $\mathcal{L}_{F_2}$ that includes the interaction between a lepton or quark pair and a Higgs field and a gauge field and is demonstrated as follows: 

\begin{eqnarray}\label{lf2}	
      \begin{split}
  \mathcal{L}_{F_2} = &
  + \frac{4 g \bar c_{eW}}{m_W^2}   y_{e} \Phi({\bar L}_L T_{2k}) \gamma^{\mu\nu} e_R  W_{\mu\nu}^k 
   + \frac{2 g' \bar c_{eB}}{m_W^2}  y_{e} \Phi \bar{ L}_L \gamma^{\mu\nu} e_R B_{\mu\nu}
   - \frac{4 g_s \bar{c}_{uG}}{m_W^2} y_u \Phi^\dag \bar{Q}_L \gamma^{\mu\nu} T_a u_R G_{\mu\nu}^a \\
   &
    - \frac{4 g \bar{c}_{uW}}{m_W^2}   y_u \Phi^\dag (\bar {Q}_L T_{2k}) \gamma^{\mu\nu} u_R  W_{\mu\nu}^k
     - \frac{2 g'\bar{c}_{uB}}{m_W^2}  y_u \Phi^\dag \bar {Q}_L \gamma^{\mu\nu} u_R B_{\mu\nu}
      + \frac{4 g_s \bar c_{dG}}{m_W^2} y_d \Phi \bar{Q}_L \gamma^{\mu\nu} T_a d_R G_{\mu\nu}^a \\
      &
       + \frac{4 g\ \bar c_{dW}}{m_W^2} y_d \Phi(\bar{Q}_L T_{2k}) \gamma^{\mu\nu} d_R  \ W_{\mu\nu}^k
   + \frac{2 g' \bar{c}_{dB}}{m_W^2}  y_d \Phi \bar{Q}_L \gamma^{\mu\nu} d_R B_{\mu\nu} +  {\rm h.c.}.  
 \end{split}
\end{eqnarray}
In this equation, $\gamma^{\mu\nu}$ is
\begin{eqnarray}
  \gamma^{\mu\nu} = \frac{i}{4} \Big[\gamma^\mu,\gamma^\nu\Big] \ 
\end{eqnarray}
And the last term in this effective Lagrangian contains those terms that do not directly include Higgs field but, rather include gauge fields. This term is represented as 
\begin{eqnarray}\label{lg1}
	\begin{split}
  \mathcal{L}_{G} = &\
   \frac{\bar{c}_{2G}}{m_W^2} D^\mu G^a_{\mu\nu} D_\rho G_a^{\rho\nu}
  + \frac{\bar{c}_{2B}}{m_W^2} \partial^\mu B_{\mu\nu} \partial_\rho B^{\rho\nu}
  + \frac{g_s^3 \bar{c}_{3G}}{m_W^2} f_{abc} G^a_{\mu\nu} G_\rho^{\nu b} G^{\rho\mu c} \\
  &\
   + \frac{g^3 \bar{c}_{3W}}{m_W^2} \epsilon_{ijk} W_{\mu\nu}^i W_\rho^{\nu j} W^{\rho\mu k}
    + \frac{\bar{c}_{2W}}{m_W^2} D^\mu W^k_{\mu\nu} D_{\rho} W^{\rho\nu}_k 
   \end{split}
\end{eqnarray}

As the oblique parameters $S$ and $T$ can be measured precisely via electroweak precision test (EWPT) and since these parameters are directly related to  $\bar c_{B} + \bar c_{W}$ and $\bar c_{T}$, the number of free Lagrangian parameters can be reduced. Regarding the corresponding constraints derived from these tests,  $\bar c_{B} + \bar c_{W}$ and $\bar c_{T}$ have very small values ~\cite{Contino:2013kra,Ellis:2014jta}.\\
After electroweak symmetry breaking, this effective Lagrangian can be expressed in the mass basis.
%After electroweak symmetry breaking, when the particles are charged, it is time to go to the mass basis. 
In this paper, effects of all Wilson coefficients on the $t\bar{t}b\bar{b}$ event production cross-section are investigated. By assigning the value of 0.1 to all parameters individually, it was concluded that the coefficients in Table.\ref{tab:Wilcoef}, which have been represented on the different sectors of effective Lagrangian, can change the $t\bar{t}b\bar{b}$ events production cross-section compared to that of the SM. 
% %%%%%%%%%%%   Table 1   %%%%%%%%%%%%%%%%%%
 %%%%%%%%%%%%%%%%%%%%%%%%%%%%%%%%%%%
 
 \begin{table}[!ht]
 	\caption{\label{tab:Wilcoef} Wilson coefficients affecting $t\bar{t}b\bar{b}$ events production cross-section.}
 	\begin{center}
 		\begin{tabular}{|c|c|}
 			\hline
 			term &  Wilson coefficient 
 			\\
 			\hline 
 			\rule{0pt}{13pt}
 			${\cal L}_{SILH}$ &   $\bar{c}_{H},\bar{c}_{u},\bar{c}_{d},\bar{c}_{\gamma},\bar{c}_{g},\bar{c}_{HB},\bar{c}_{HW}$ 
 			\\
 	 		\hline
			 \rule{0pt}{13pt}
 			${\cal L}_{CP}$ &     $\tilde{c}_{\gamma},\tilde{c}_{g},\tilde{c}_{HB},\tilde{c}_{HW}$
 			\\
 			\hline
 			\rule{0pt}{13pt}
 			${\cal L}_{F_{1}}$ &  $\bar{c}_{HQ},\bar{c'}_{HQ},\bar{c}_{Hu},\bar{c}_{Hd},\bar{c}_{Hud}$  
 			\\
 			\hline
 			\rule{0pt}{13pt}
 			${\cal L}_{F_{2}}$ & $\bar{c}_{dB},\bar{c}_{dW},\bar{c}_{uB},\bar{c}_{uW},\bar{c}_{uG},\bar{c}_{dG}$
 			\\
 			\hline
 			\end{tabular}
 	\end{center}
 \end{table}
 %%%%%%%%%%%%%%%%%%%%%%%%%%%%%%%%%%%%%%%%%%%%%%%%%%%%%%%%%%%%%%%%%%%%%%%%%%%
 
\begin{figure}[hbpt]
          \begin{center}
            \subfigure[]{
            \includegraphics[width=4.5cm]{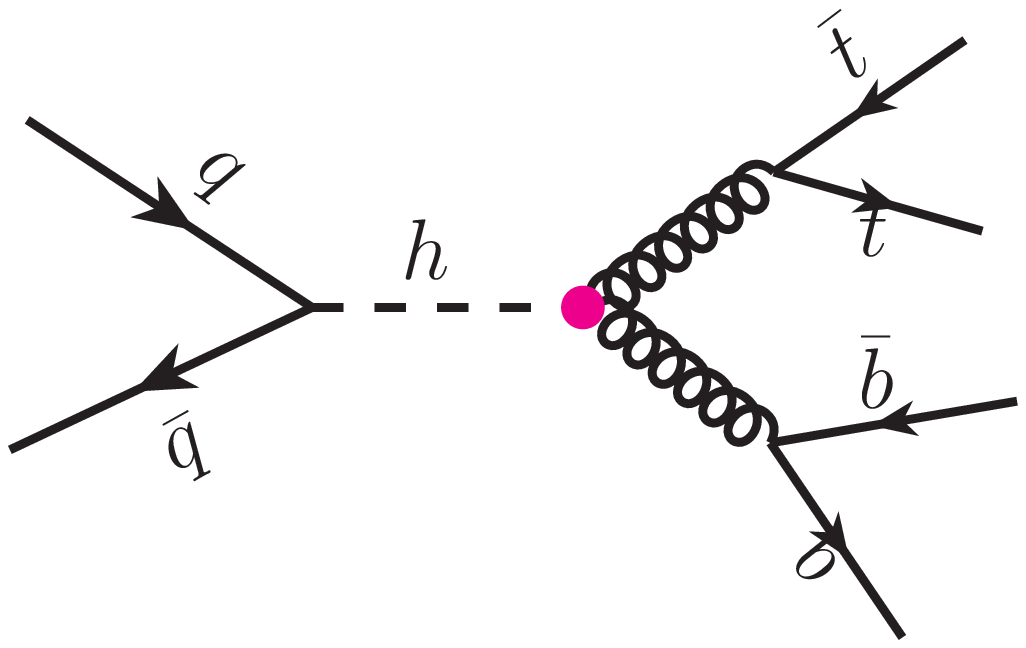}\label{FEYN1-a}}
                \hspace{0.5cm}
            \subfigure[]{
             \includegraphics[width=4.5cm]{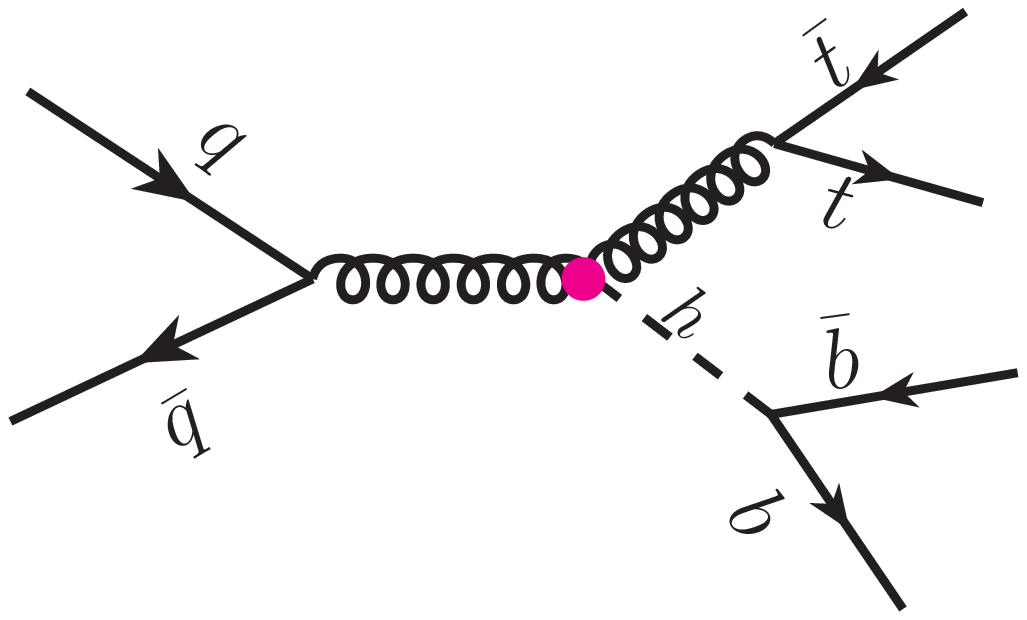}\label{FEYN1-b}}
                \hspace{5cm}
                \hfill
             \subfigure[]{
             \includegraphics[width=4.5cm]{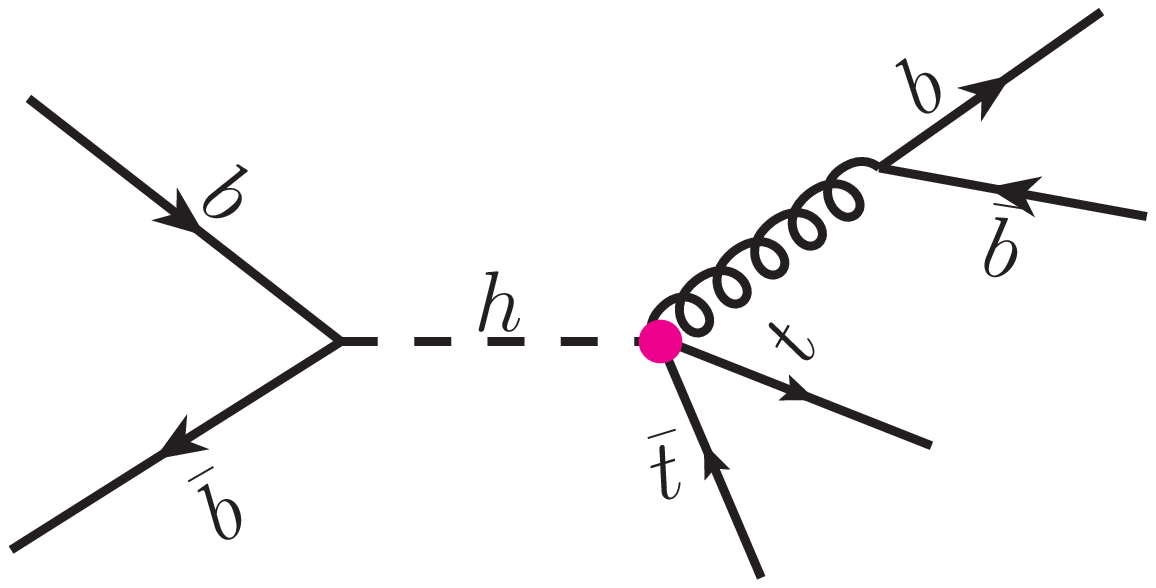}\label{FEYN1-c}}
                \hspace{0.5cm}
                 \subfigure[]{
             \includegraphics[width=4.5cm]{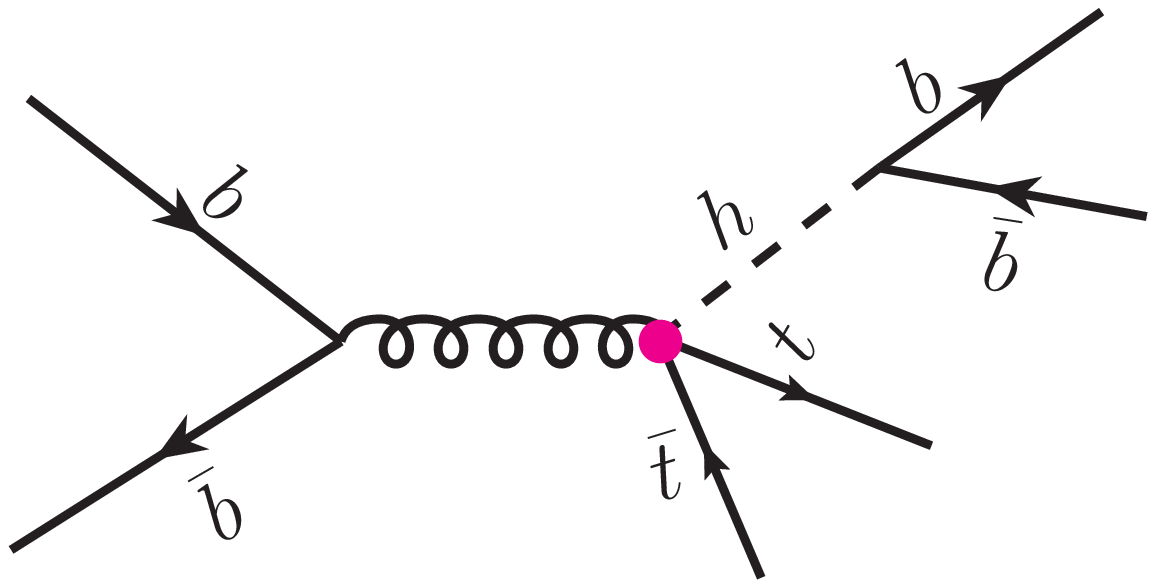}\label{FEYN1-d}}
                \hspace{0.5cm}
                \hfill
            \caption{\label{FEYN1} Feynman diagrams of $t\bar{t}b\bar{b}$ events production containing new interactions of gluon boson and Higgs boson.  }
          \end{center}
        \end{figure}

\begin{figure}[t!]
  \centering
   \subfigure{\includegraphics[width=4.5cm]{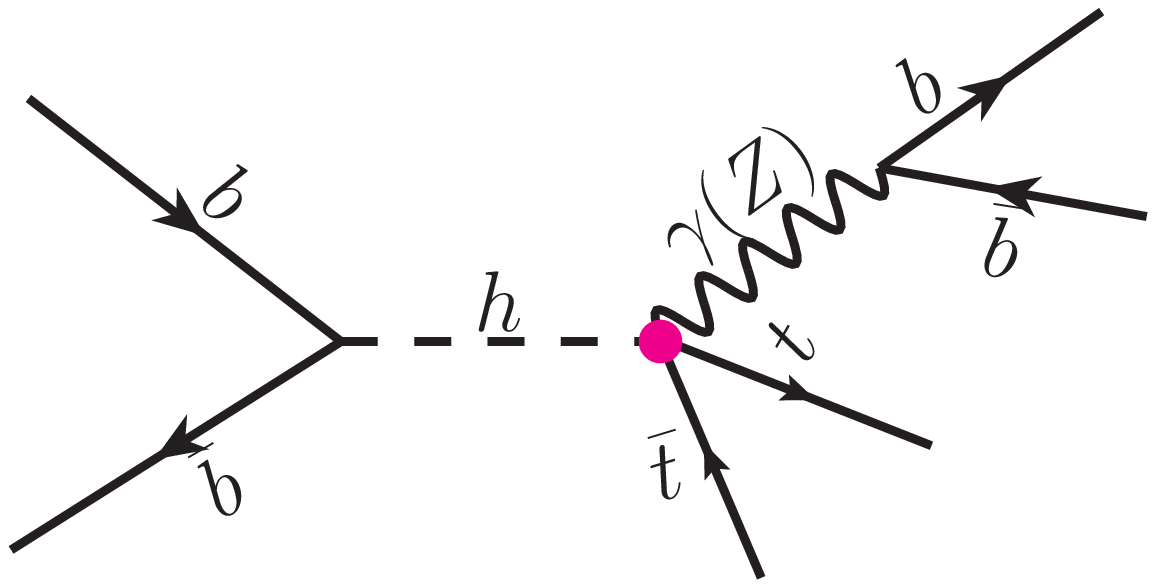}}
    \subfigure{\includegraphics[width=4.5cm]{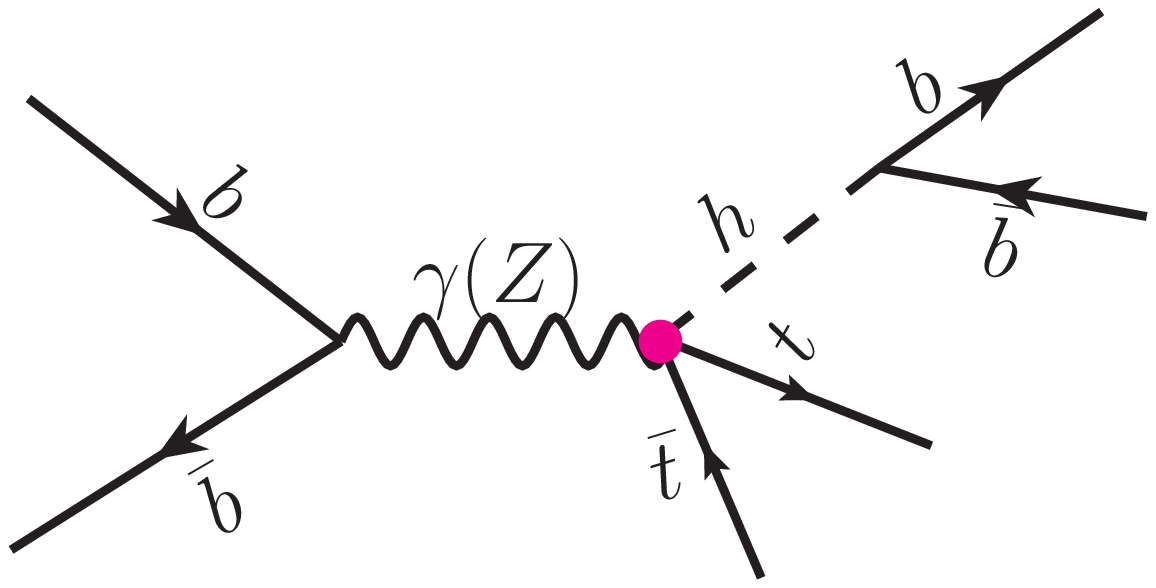}}
     \subfigure{\includegraphics[width=4.5cm]{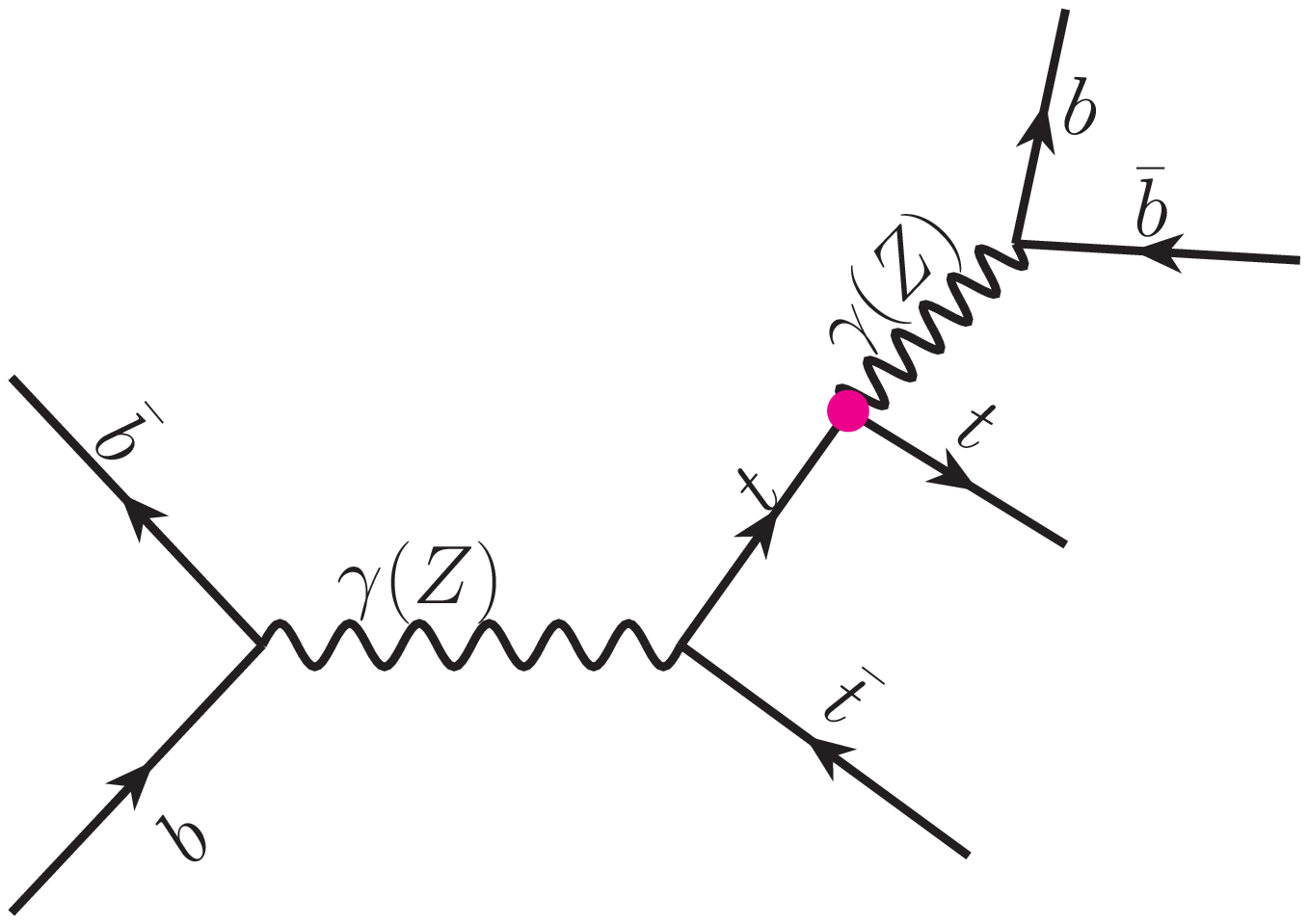}}
     \subfigure{\includegraphics[width=4.5cm]{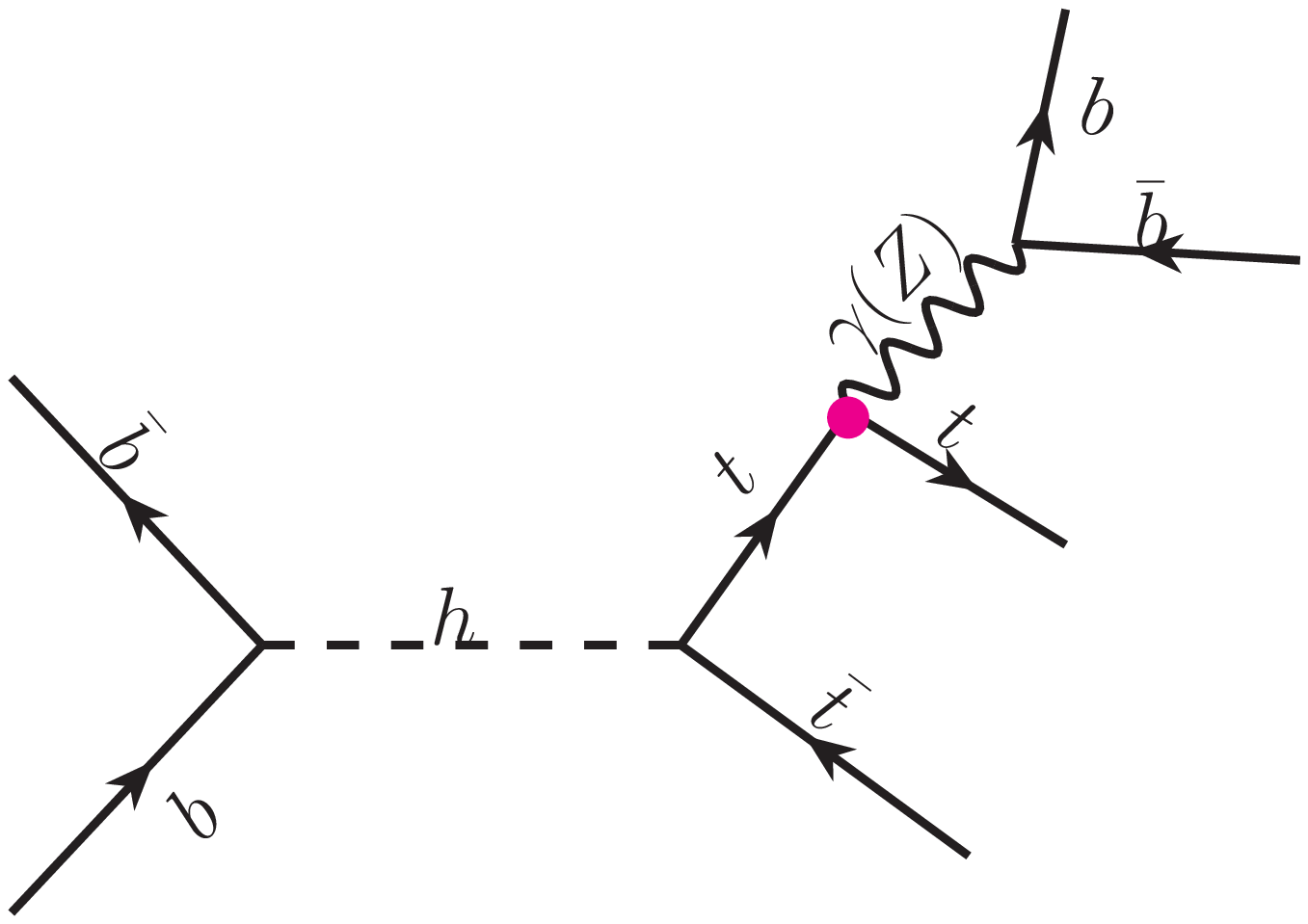}}
   \subfigure{\includegraphics[width=4.5cm]{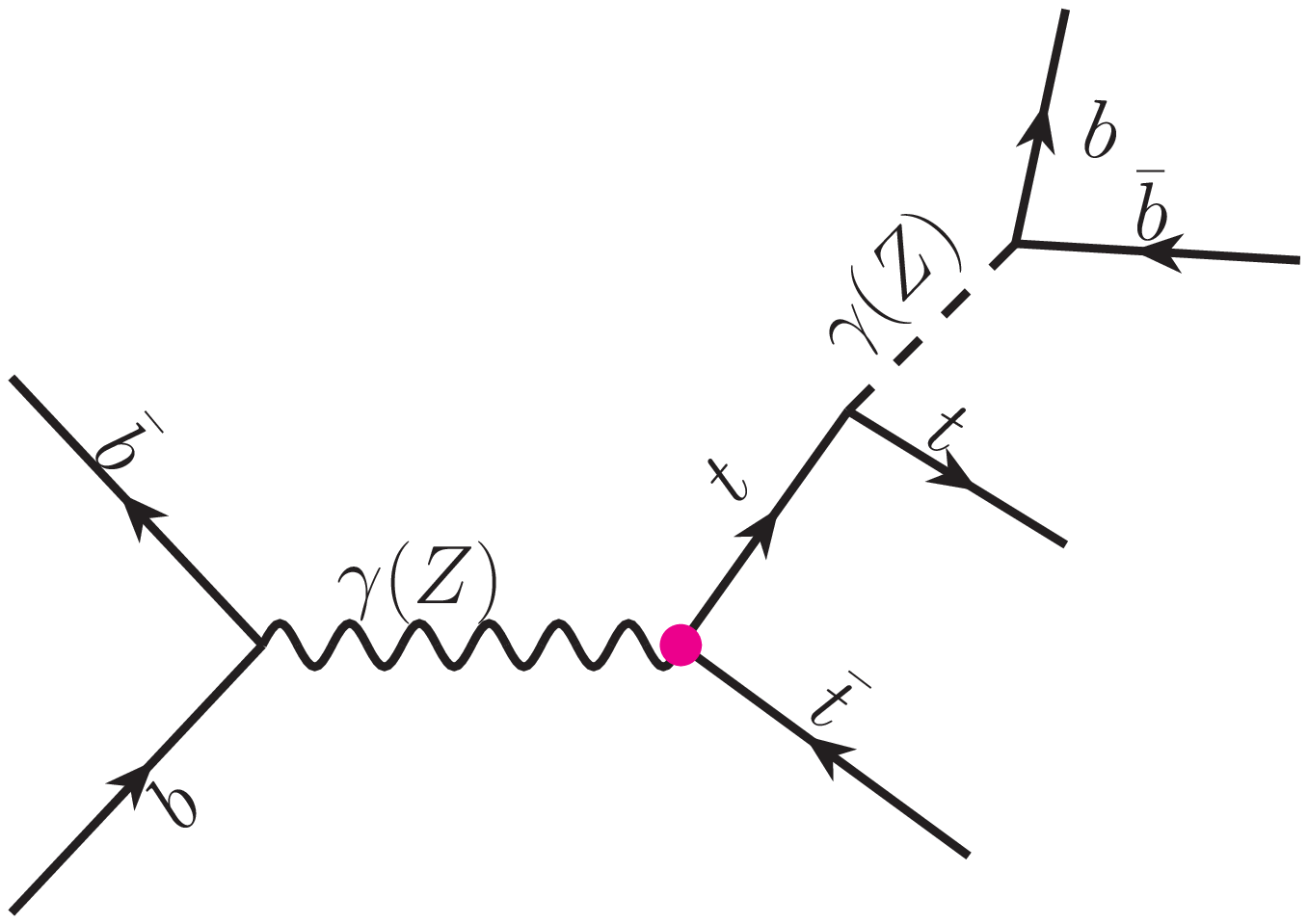}}
   \subfigure{\includegraphics[height=2.5cm]{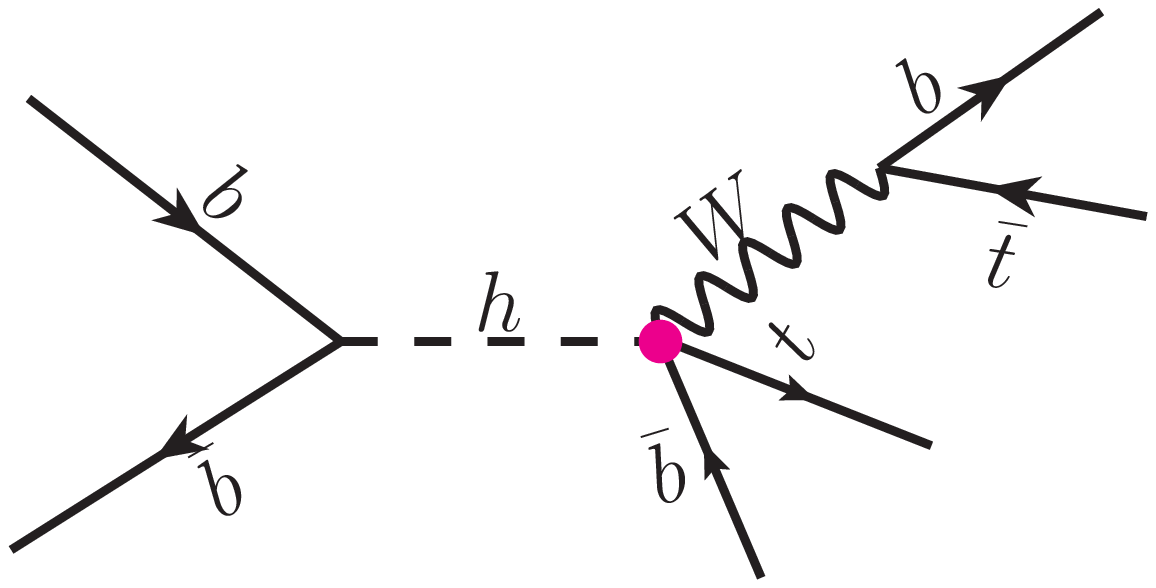}}
     
      \caption{New Feynman diagrams of $t\bar{t}b\bar{b}$ events production derived from those terms containing the interaction between a quark pair, a Higgs boson, and a gauge boson $\gamma$ or Z or W.}
\label{FEYN2}
\end{figure}

 Among all the Wilson coefficients, only four, namely $\bar{c}_{g},\tilde{c}_{g},\bar{c}_{uG},\bar{c}_{uW} $, could change the $t\bar{t}b\bar{b}$ events production cross-section at 13TeV by more than $4\%$. 
$\bar{c}_{g}$ and $\tilde{c}_{g}$ represent the coefficients of the terms $\Phi^\dag \Phi G_{\mu\nu}^a {G}^{\mu\nu}_a$ and $\Phi^\dag \Phi G_{\mu\nu}^a {\widetilde G}^{\mu\nu}_a$, respectively. After the electroweak symmetry breaking, these terms can  modify the Feynman rules of SM, in diagrams with three or four gluons. Besides, they can add some new Feynman rules containing gluon fields associated with one or two Higgs fields to the SM Feynman rules. These new vertices that contain Higgs particles and gluon fields can directly affect the Higgs particle production events at the LHC.
Figures \ref{FEYN1-a} and \ref{FEYN1-b} show two samples of the new Feynman diagrams, which are related to the production of $t\bar{t}b\bar{b}$ events resulting from these two new reactions. \\
Both of couplings $\bar{c}_{uG}$ and $\bar{c}_{uW} $ are located in the Lagrangian section ${\cal L}_{F_{2}}$ and represent the interactions between a quark pair, a Higgs field, and a gauge field. The interesting point about the two-fermion operators is that, these terms are quite similar in the SILH and Warsaw bases \cite{Ferreira:2016jea}. Generally speaking, these terms are recognized as dipole interactions, which are physical observables; thus, their corresponding Wilson coefficients would be basis independent. Figures \ref{FEYN1-c}, \ref{FEYN1-d}, and \ref{FEYN2} exemplify the new Feynman diagrams that have been derived from the coefficient terms $\bar{c}_{uW}$ and $\bar{c}_{uG}$, respectively.\\
%After the electroweak symmetry breaking, these coefficients can contribute to the modification of the vertices provided in Tables \ref{Tab1}, \ref{Tab2}, and \ref{Tab3} in the appendix. Subsequent to the 
After the electroweak symmetry breaking, the effective Lagrangian with the greatest effect on $t\bar{t}b\bar{b}$ events production will be as follows \cite{PAlloul:2013naa}: 
\begin{eqnarray}\label{massbais}
\begin{split}
	&\
  - \frac{1}{4} \tilde g_{hgg} G^a_{\mu\nu} \tilde G^{\mu\nu} h-\frac{1}{8} \tilde g_{hhgg} G^a_{\mu\nu} \tilde G_a^{\mu\nu} h^2 
   -[{\bar u} \gamma^\mu \Big[ g_{hwud}^{(L)} P_L +  g_{hwud}^{(R)} P_R \Big] d W_\mu h] \\
 &\
 - \frac{1}{4} g_{hgg} G^a_{\mu\nu} G_a^{\mu\nu} h- \frac{1}{8} g_{hhgg} G^a_{\mu\nu} G_a^{\mu\nu} h^2
 -[g_{h\gamma uu}^{(\partial)} [ {\bar u} \gamma^{\mu\nu} P_R u]] F_{\mu\nu} h \\
     &\
    -[g_{hzuu}^{(\partial)} [ {\bar u} \gamma^{\mu\nu} P_R u ]]  Z_{\mu\nu} h 
    -[g_{hguu}^{(\partial)} [ {\bar u} T_a \gamma^{\mu\nu} P_R u ]] G_{\mu\nu}^a h + {\rm h.c.} 
  \end{split}
    \end{eqnarray}
Following the occurrence of electroweak symmetry breaking, the coefficients $\bar{c}_{g}$ and $\tilde{c}_{g}$ exist in coefficients $g_{hgg},g_{hhgg}$ and $\tilde g_{hgg},\tilde g_{hhgg}$, respectively. $\bar{c}_{uW}$  appears in $g_{hwud}^{(L)},g_{hwud}^{(R)},g_{hzuu}^{(\partial)},g_{h\gamma uu}^{(\partial)}$ and the coefficient $\bar{c}_{uG}$ in $g_{hguu}^{(\partial)}$.\\
%To exemplify, the diagrams that are capable of applying further changes in cross-section of the $t\bar{t}b\bar{b}$ events production are demonstrated in Figures (2) and (3). Figure (3) represents the new diagrams that are derived from the terms containing the interaction between a quark pair, a Higgs boson, and a gauge boson $\gamma$ or Z or W. Preceded by electroweak symmetry breaking, the vertices include the coefficients $g_{hwud}^{(L)},g_{hwud}^{(R)},g_{hzuu}^{(\partial)},g_{h\gamma uu}^{(\partial)}$. Figure (3) shows the Feynman diagrams of $t\bar{t}b\bar{b}$ events production containing the new interactions of gluons and Higgs boson. After the electroweak symmetry breaking, these figures would include the coefficients $g_{hguu}^{(\partial)},g_{hgdd}^{(\partial)},g_{hgg},\tilde{g}_{hgg}$.  Figure-3: Feynman diagrams of $t\bar{t}b\bar{b}$ events production containing new interactions of gluon bosons and Higgs boson 

 %
%%%%%%%%%%%%%%%%%%%%%%%%%%%%%%%%%%%%%%%%%%%%%%%%%%%%%%%%%%%%%%%%%%%%%%
\section{Simulation and analysis methods}\label{sec:analysis}
%%%%%%%%%%%%%%%%%%%%%%%%%%%%%%%%%%%%%%%%%%%%%%%%%%%%%%%%%%%%%%%%%%%%%%
In this section, we deal with simulating the $t\bar{t}b\bar{b}$ events caused by the collision of the protons with center-of-mass energy of 13TeV. Also, the results obtained from CMS group's study on di-leptonic $t\bar{t}b\bar{b}$ events are used to investigate the effective Lagrangian coefficients. For this purpose, first, the entire effective Lagrangian should be imported to {\tt FeynRules}~\cite{PChristensen:2008py,PAlloul:2013bka} in order to create the model as a UFO (Universal Feynrules Output)~\cite{PDegrande:2011ua,PAlloul:2013naa}. Then, the model-corresponding UFO file should be inserted in the {\tt MadGraph5-aMC@NLO}~\cite{PAlwall:2011uj,PAlwall:2014hca}. 
%Monte Carlo is capable of producing the events related to $ p p \to t \bar{t} b \bar{b}$ as well as calculating the cross-section of these events.
With the help of MadGraph cross-sections were calculated and Monte Carlo events were generated. For the purpose of considering one-loop modifications, K-factor is equal to 1.8~\cite{Bevilacqua:2009zn}. In this analysis, both top quarks exhibited leptonic decay. This was performed using {\tt MadSpin} module~\cite{Artoisenet:2012st,Frixione:2007zp} package. The parton distribution function CTEQ6L1 was considered for the proton. {\tt Pythia 8}~\cite{Sjostrand:2003wg,Sjostrand:2007gs} package is utilized to perform  fragmentation, hadronisation, initial- and final-state parton showers. Also, anti$k_T$~\cite{Soyez:2008pq}, algorithm and {\tt FastJet3.2.0}~\cite{Cacciari:2011ma} software were employed to cluster the jets, for the purpose of which the value of $\Delta$R was considered equal to 0.4. \\
The CMS group has reported the $t\bar{t}b\bar{b}$ events cross-section measurement for two different regions: visible space and full phase space. Visible phase space refers to a region where all particles, except for neutrinos, including leptons and jets existing within the event are required to be accessible in the detector. The conditions considered in this paper are the same as those applied by CMS group on the events. Accordingly, pseudo-rapidity of the jets and leptons were considered within the range of $|\eta|<2.5$ and $|\eta|<2.4$ , respectively and minimum transverse momentum of them  were 20GeV/C. Also, the minimum number of b-jets and leptons was considered equal to 2 for each event. The bottom-jet (b-jet) tagging efficiency was $70\%$, but the mis-tagging probability was shown to be $10\%$ for the misidentification of charm-jets (c-jets) as b-jets and $1\%$ for other light quarks. The major backgrounds of the $t\bar{t}b\bar{b}$ signal events include Z+jets events, single top quark events, $t\bar{t}$ events associated with a gauge boson, $t\bar{t}$ associated with a light quark pair, as well as $c\bar{c}$ and $t\bar{t}b$ events associated with a light quark. 
% ($ t\bar{t}b\bar{b} \to b b W^{-} W^{+}\bar{b} \bar{b} \to b b l^{-} \bar{\nu} l^{+} \nu \bar{b} \bar{b}$) 
An investigation of the CMS group's study at the energy of 13TeV seems to suffice for the intended discussion on backgrounds in the present paper \cite{CMS:2016tlo}.\\
 According to the CMS group's analysis results, the cross-section that has been measured for $t\bar{t}b\bar{b}$ events in the visible phase space is equal to $0.088\pm0.012(stat)\pm0.029(syst)$pb. The SM prediction up to the NLO order for the same visible phase space is equal to $0.07\pm0.009$pb. Accordingly, the measured values exhibited the deviation of 2$\sigma$ from the SM predicted value \cite{CMS:2016tlo}.\\
By simulating the $ p p \to t \bar{t} b \bar{b}$ events for a center-of-mass in the presence of the above-mentioned cuts, the signal strength dependency relative to $\bar{c}_{g},{\tilde{c}}_{g},\bar{c}_{uG},\bar{c}_{uW}$ was determined as follows: \\
\begin{eqnarray}
  \begin{split}
  \mu_{\bar{c}_{uG}}=\frac{\sigma_{\bar{c}_{uG}}}{\sigma_{SM}}=1+195.0\times\bar{c}_{uG}+22700\times{\bar{c}}^{2}_{uG}\,\\
  \mu_{\bar{c}_{uW}}=\frac{\sigma_{\bar{c}_{uW}}}{\sigma_{SM}}=1+0.197\times\bar{c}_{uW}+55.36\times{\bar{c}}^{2}_{uW}\,\\
  \mu_{\bar{c}_{g}}=\frac{\sigma_{\bar{c}_{g}}}{\sigma_{SM}}=1+112.86\times\bar{c}_{g}+7628.57\times{\bar{c}}^{2}_{g}\,\\
  \mu_{{\tilde{c}}_{g}}=\frac{\sigma_{{\tilde{c}}_{g}}}{\sigma_{SM}}=1+22.143\times\bar{\tilde{c}}_{g}+
  4785.71\times{\tilde{c}}^{2}_{g}\,\\
  \end{split}
\end{eqnarray}

%
%%%%%%%%%%%%%%%%%%%%%%%%%%%%%%%%%%%%%%%%%%%%%%%%%%%%%%%%%%%%%%%%%%%%%%
\section{Results}\label{sec:results}
%%%%%%%%%%%%%%%%%%%%%%%%%%%%%%%%%%%%%%%%%%%%%%%%%%%%%%%%%%%%%%%%%%%%%%
%
In this section, the analysis results are shown and a comparison with other studies are presented. Since the cross-section measured for $ t \bar{t} b \bar{b}$ event production at 13TeV at the LHC shows deviaton from the SM prediction\cite{CMS:2016tlo}, it can serve as a hint for new physics, through which the Wilson coefficients in effective Lagrangian can be constrained at the confidence level of $95\%$C.L. These constraints are provided in the following Table.\ref{tab:constraints23}: 
%%%%%%%%%%%%%%%%%%%%%%%%%%%%%%%%%%%%%%
\begin{table*}[!htb]
\begin{center}
\begin{tabular}{ | c | c | }
% \\[0.1cm]
\hline \rule{0pt}{13pt}
 $\bar{c}_{uG}$ & $[\,-3.6\times10^{-5}, \,3.6\times10^{-5} \, ]$   \\ \hline \rule{0pt}{13pt}
 $\bar{c}_{uW}$ & $[\,-9.8\times10^{-3}, \,1.3\times10^{-2}\, ]$  \\ \hline \rule{0pt}{13pt}
$\bar{c}_{g}$ & $[\,-6.3\times10^{-5}, \,6.2\times10^{-5}\, ]$  \\  \hline \rule{0pt}{13pt}
 $\tilde{c}_{g}$ & $[\,-3.4\times10^{-4}, \,3.0\times10^{-4}\, ]$  \\ \hline 
\end{tabular}
\caption{Constraints obtained at $95\%$ confidence level for $2.3~\text{fb}^{-1}$ data of $p p \rightarrow t \bar{t}b\bar{b}$ at 13TeV.
\label{tab:constraints23}}
 \end{center}
\end{table*}
%%%%%%%%%%%%%%%%%%%%%%%%%%%%%%%%%%%%%%
\\
%Now, we are going to investigate the results for the case with greater amounts of data. 
In this step, we study how LHC sensitivity improves with large luminosity. Some of the systematic errors considered in CMS group's study include background, JER, JES, Pileup, Lepton trigger/identification, b-tag, and modelling, among which the error resulted from b-tagging had the highest contribution to the systematic error\cite{CMS:2016tlo}. \\
Generally, some types of systematic errors depend on the amount of data, but their dependencies are not similar to that of the statistical error; thus, it cannot be easily predicted. For the systematic error, it is only possible to assume some value for improving the systematic errors by increasing the amount of data. \\
In this paper, the Systematic and statistical errors can be extrapolated from two approaches, in the first of which only the statistical error is reduced with an increase in data, but the systematic error rate remains unchanged. In such a case, the predicted statistical error for measuring the cross-section of $p p \rightarrow t \bar{t}b\bar{b}$ events production at 13TeV for 40fb$^{-1}$ and 3000fb$^{-1}$, compared to the error for 2.3fb$^{-1}$, was equal to 23.9$\%$ and 2.7$\%$, respectively. In the first approach, since the majority of the errors is due to the systematic errors, the 40fb$^{-1}$ and 3000fb$^{-1}$ data yield similar constraints presented in the first column of Table\ref{tab:projectconstrants}. Furthermore, due to the greater systematic error, the constraints obtained with larger data do not greatly differ from those obtained for 2.3fb$^{-1}$ integrated luminosity. However, in the second approach, both statistical and systematic errors undergo changes. 
%%%%%%%%%%%%%%%%%%%%%%%%%%%%%%%%%%%%%%
\begin{table*}[htb]
%[tb]
\begin{tabular}{  c | c | c |c | c |}
% \\[0.1cm]
%\hline
  &1st approach & 2nd approach(40 fb$^{-1}$) & 2nd approach(3000 fb$^{-1}$) & LHC\cite{Englert:2015hrx,Ferreira:2016jea,Sirunyan:2017uzs} \\
\hline \hline \rule{0pt}{13pt}
 $\bar{c}_{uG}\times10^{5}$ & $[\,-3.40, \,3.30 \, ]$ & $[\,-0.87, \,0.87 \, ]$  & $[\,-0.10, \,0.10 \, ]$  & $[\, -700, \,200 \, ]$\cite{Sirunyan:2017uzs}    \\ \hline \rule{0pt}{13pt}
$\bar{c}_{uW}\times10^{3}$ & $[\,-9.40, \,13 \, ]$ &$[\,-4.20, \,7.40 \, ]$& $[\,-0.86, \,4.10 \, ]$ &  $[\, -14.0, \,14.0 \, ]$ \cite{Sirunyan:2017uzs} \\ \hline \rule{0pt}{13pt}
$\bar{c}_{g}\times10^{5}$ & $[\,-5.80, \,5.80 \, ]$ &$[\,-1.50, \,1.50 \, ]$&  $[\,-0.17, \,0.17 \, ]$&$[\, -8.2, \,>10.0 \, ]$\cite{Englert:2015hrx} \\ \hline \rule{0pt}{13pt}
$\tilde{c}_{g}\times10^{5}$ &$[\,-32.0, \,28.0 \, ]$ &$[\,-7.80, \,7.60 \, ]$&  $[\,-0.88, \,0.89 \, ]$&$[\, -6.00, \,6.00 \, ]$\cite{Ferreira:2016jea}\\ \hline
\end{tabular}
\caption{The predicted constraints on wilson coefficients for the LHC at 13 TeV for integrated luminosities of $40~\text{fb}^{-1}$ and $3000~\text{fb}^{-1}$ in two approaches. The fifth column are the results of previous studies.
\label{tab:projectconstrants}}
\end{table*}
%%%%%%%%%%%%%%%%%%%%%%%%%%%%%%%%%%%%%%
In the second approach, it is assumed that with an increase in the data at 13TeV, the systematic error will be reduced as much as the statistical error. Regarding such an assumption that is not out of reach, for the confidence level of $95\%$, the effective Wilson coefficients were constrained for 40fb$^{-1}$ and 3000fb$^{-1}$, which are presented in the second and third columns of Table.\ref{tab:projectconstrants}. These constraints exhibit a smaller window in the parameter space compared to the constrained obtained for 2.3fb$^{-1}$ integrated luminosity. \\
% In this approach stronger constraints are found relative to the first approach.\\
In the second approach, $\bar{c}_{g}$ is constrained to $[\, -1.7\times10^{-6}, \,1.7\times10^{-6} \, ]$ from the cross-section of $t\bar{t}b\bar{b}$ events production with 3000fb$^{-1}$ data. In \cite{Englert:2015hrx}, signal strength and transverse momentum of the Higgs boson in $p p \to H,H+j,H+2j,\bar{t}tH$ events were used, which yielded constrained $\bar{c}_{g}$. The best constraint at the confidence level of $95\%$C.L for proton-proton collision at center-of-mass energy 14TeV and data amount of 3000fb$^{-1}$ was $[\, -8.2\times10^{-5}, \,>10^{-4} \, ]$. Therefore, the constraint obtained is an improvement over previous studies by about two orders of magnitude. \\
The constraint obtained in the present study for ${\tilde{c}}_{g}$ coefficient for data amount of 3000fb$^{-1}$ in the second approach is $[\, -8.8\times10^{-6}, \,8.9\times10^{-6} \, ]$. In \cite{Ferreira:2016jea}, the authors investigated the CP-violating Wilson coefficients at LHC, one of which was ${\tilde{c}}_{g}$. The constraint obtained on the Wilson coefficients for the LHC at center of mass energy 13TeV for integrated luminosity 3000fb$^{-1}$ was $|{\tilde{c}}_{g}|<6\times10^{-5}$, which is a milder constraint than the one, we obtained in this paper. In \cite{Ferreira:2016jea}, 
%implies stronger constraint obtained in the present study. In their work, they yield the sensitivity of 
the production of $g g \to h \to \gamma \gamma$ events to Wilson coefficients ${\tilde{c}}_{g}$ and ${\tilde{c}}_{\gamma}$ :  
\begin{eqnarray}
   \mu^{g g \to h \to \gamma \gamma}=1.0+2\times10^{5}{\tilde{c}}_{g}^{2}-1.5\times10^{4}{\tilde{c}}_{g}{\tilde{c}}_{\gamma}
  +2\times10^{7}{\tilde{c}}_{\gamma}^{2}
\end{eqnarray}
%for the signal strength dependency of the production of $g g \to h \to \gamma \gamma$ events on the Wilson coefficients ${\tilde{c}}_{g}$ and ${\tilde{c}}_{\gamma}$ in the case of 13TeV center-of-mass energy. 
The use of signal strength of $g g \to h \to \gamma \gamma$ events along with the $t \bar{t}b\bar{b}$ events production cross-section can help constraining ${\tilde{c}}_{g}$  
%more than the one in \cite{Ferreira:2016jea}, which would yield the constraint
to $[\, -2.3\times10^{-5}, \,2.3\times10^{-5} \, ]$ for the data amount of 3000fb$^{-1}$, which is an improvement over \cite{Ferreira:2016jea}.\\ 
In \cite{Sirunyan:2017uzs}, the authors studied the $t\bar tW$ and $t\bar tZ$ events at proton-proton collision in the center-of-mass energy of 13TeV for 35.9fb$^{-1}$ data. According to the results, the coefficients $\bar{c}_{uG}$ and $\bar{c}_{uW}$ affected the production of $t\bar tW$ and $t\bar tZ$ events. Based on the results of their study, the two coefficients were constrained to the ranges of $[\, -0.007, \,0.002 \, ]$ and $[\, -0.014, \,0.014 \, ]$, respectively. Comparing these constraints with the values reported in Table (3) for 40fb$^{-1}$ in the second approach reveals that the cross-section of the $t\bar{t}b\bar{b}$ events production can put stronger constraint on $\bar{c}_{uG}$ and $\bar{c}_{uW}$. In \cite{Sirunyan:2017uzs}, the signal strength dependency of the $t\bar tW$ and $t\bar tZ$
 events on $\bar{c}_{uG}$ and $\bar{c}_{uW}$ coefficients at the LHC in the case of the center-of-mass energy of 13TeV can be expressed as follows: 
\begin{eqnarray}
 \begin{split}
  \mu^{pp\to t\bar{t}W}_{\bar{c}_{uW}}=0.999+2.43\bar{c}_{uW}+628.3\bar{c}_{uW}^{2},
  \mu^{pp\to t\bar{t}W}_{\bar{c}_{uG}}=1.01+70.77\bar{c}_{uG}+1740.2\bar{c}_{uG}^{2},\\
  \mu^{pp\to t\bar{t}Z}_{\bar{c}_{uW}}=0.985+0.458\bar{c}_{uW}+1994\bar{c}_{uW}^{2},
  \mu^{pp\to t\bar{t}Z}_{\bar{c}_{uG}}=1.03+97.9\bar{c}_{uG}+14705.9\bar{c}_{uG}^{2},
   \end{split}
\end{eqnarray}
Using the cross-sections of three events $t\bar{t}b\bar{b}$, $t\bar tW$ and $t\bar tZ$ at the LHC in the center of mass energy 13TeV, the two coefficients $\bar{c}_{uG}$ and $\bar{c}_{uW}$ can be constrained. Therefore, in the integrated luminosity of 35.9fb$^{-1}$ , the coefficients $\bar{c}_{uG}$ and $\bar{c}_{uW}$ can be constrained at the $95\%$ confidence level  as $[\, -7.8\times10^{-5}, \,7.3\times10^{-5} \, ]$ and $[\, -7.5\times10^{-3}, \,7.2\times10^{-3} \, ]$, respectively. These constraints are improved by 2 and 1 orders of magnitude compared to those mentioned in \cite{Sirunyan:2017uzs}. \\
Accordingly, production of the $t\bar{t}b\bar{b}$ events at the LHC is very sensitive to coefficients $\bar{c}_{g},{\tilde{c}}_{g},\bar{c}_{uW},\bar{c}_{uG}$. Thus, precise measurement of this process at the LHC can greatly help investigate physics beyond SM. 

\section{Summary and Conclusion}
\label{summ}
So far, we have not found any unambiguous sign of new physics. Therefore, the new physics scale seems to be well separated from the electroweak scale. Furthermore, in numerous new models that contain new particles, such a heavy degrees of freedom can be integrated out. This evidence has motivated us to use the model-independent effective theory approach. If we constrain ourselves to SM gauge symmetries, Lorentz invariance, as well as Lepton and Baryon number conservation, the first order corrections to the SM occur at dimension six; even though the contribution of dimension six operators are reduced by the second power of the new physics scale.\\
The present paper focused on the effect of all dimension six operators of the SM effective field theory on SILH basis on the production of $t\bar{t}b\bar{b}$ events at the LHC. Once the Wilson coefficients affecting the production of $t\bar{t}b\bar{b}$ events were found, the focus was put on those coefficients that were capable to change the cross-section of production of these events at the center-of-mass energy of 13TeV by more than $4\%$ compared to the SM predictions. For this purpose, these coefficients were constrained using the most recent measurement results obtained for cross-section of the production of $t\bar{t}b\bar{b}$ events in CMS experiment. The obtained results indicated that the production of $t\bar{t}b\bar{b}$ events at LHC allows constraining the $\bar{c}_{g}$,$\tilde{c}_{g}$,$\bar{c}_{uG}$ up to $10^{-6}$ and $\bar{c}_{uW} $ up to $10^{-4}$ in the integrated luminosity of 3000fb$^{-1}$. Comparing the results of the present study with other constraints obtained for these coefficients at the LHC indicates the stronger constraints of those obtained in the present work. Furthermore, it was shown that the constraints on $\tilde{c}_{g}$ coefficient obtained from $g g \to h \to \gamma \gamma$ can be improved by using these events along with the $t\bar{t}b\bar{b}$ events. As shown in this work, combining the signal strength of $t\bar tW,t\bar tZ$ and cross-section of $t\bar{t}b\bar{b}$ events to form $\chi^2$ can lead to stronger constraints on $\bar{c}_{uG}$ and $\bar{c}_{uW}$, compared to the case that only the $t\bar tW,t\bar tZ$ events are considered. On this basis, the production of $t\bar{t}b\bar{b}$ events at the LHC is very sensitive to the coefficients $\bar{c}_{g}$,$\tilde{c}_{g}$,$\bar{c}_{uG}$ and $\bar{c}_{uW} $. Therefore, precise measurement of this process at LHC can greatly help in studying physics beyond the SM. 
%
%%%%%%%%%%%%%%%%%%%%%%%%%%%%%%%%%%%%%%%%%%%%%%%%%%%%%%%%%%%%%%%%%%%%%%
\section*{Acknowledgments}
%%%%%%%%%%%%%%%%%%%%%%%%%%%%%%%%%%%%%%%%%%%%%%%%%%%%%%%%%%%%%%%%%%%%%%
H.H is particularly grateful to Mojtaba Mohammadi Najafabadi for the useful discussions and Ferdos Rezaei  and Fatemeh Elahi  for valuable comments on the manuscript. Author thanks School of Particles and Accelerators, Institute for Research in Fundamental Sciences (IPM) for financial support of this project.

\newpage
\newpage
%%%%%%%%%%%%%%%%%%%%%%%%%%%%%%%%

%%%%%%%%%%%%%%%%%%%%%%%%%%%%%%%%%%%%%%%%%%%%%%%%%%%%%%%%%%%%%%%

\end{document}